\documentclass[twocolumn,showpacs,preprintnumbers,amsmath,amssymb]{revtex4}

\usepackage{graphicx}
\usepackage{dcolumn}
\usepackage{bm}

\begin{document}

\preprint{APS/123-QED}

\title{On the genre-fication of Music: a percolation approach (long version)}

\author{R. Lambiotte}
\email{Renaud.Lambiotte@ulg.ac.be}

\author{M. Ausloos}
\email{Marcel.Ausloos@ulg.ac.be}

\affiliation{%
SUPRATECS, Universit\'e de Li\`ege, B5 Sart-Tilman, B-4000 Li\`ege, Belgium
}%

\date{09/07/2005}

\begin{abstract}
In this paper, we analyze web-downloaded data  on
people
sharing their music library. By attributing to each music group usual music genres (Rock, Pop...),
and analysing correlations between music groups of different genres with percolation-idea based methods, 
we probe the reality of these subdivisions and construct a music genre cartography, with a tree representation. We also show the diversity of music genres with Shannon entropy arguments, and discuss an alternative objective way to classify music, that is based on the complex structure of the groups audience. Finally, a link is drawn with the theory of {\em hidden variables} in complex networks. 
\end{abstract}

\pacs{89.75.Fb,  89.65.Ef, 64.60.Ak}

\maketitle

\section{Introduction}

Take a sample of people, and 
make them listen to a list of songs. If a majority of people 
should find an agreement on basic subdivisions, like {\em Rock/Jazz/Pop...}, 
a more refine description will lead to more and more disparate answers,
even contradictions. These originate from the different background, taste, music knowledge, mood 
or {\em network of acquaintances} \cite{sornette} of the listeners, i.e.  in a statistical physics description, these processes correspond to  ageing,  internal fluctuations and neighbour-neighbour interactions.
The more and more eclectic music offer,  together with the constant mixing of
old music genres into new ones make the problem still more complicated.
Even artists seem to avoid the usual classifications by refusing to enter well-defined yokes,
and prefer to characterise themselves as a unique mix-up of their old influences \cite{deus}.

Obviously, categorising music, especially into finer genres or subgenres, is not an easy task, and is
strongly subjective. This task is also complicated by the constant birth of
newly emerging styles, and by the very large number of existing sub-divisions. For instance, the genre {\em Electronic music} is divided in {\em wikipedia} \cite{wiki1} into 9 sub-genres ({\em Ambient, Breakbeat...}), each of them being divided into several subsubgenres. This categorising is becoming more and more complex in the course of time.

This paper tries to find an answer to the above problems by showing in an "objective" way the existence of music trends that allow to classify music groups, as well as the relations between the usual genres and sub-genres. 
To do so, we use web-downloaded data from the web, and define classifications based on the listening habits 
of the groups audience. Thereby, we account for the fact that music 
perception is driven both by the people who make music 
(artists, Majors...), but also 
by the people who listen to it.
Our analysis consists in characterising a large sample of individual musical habits
from a statistical physics point of view, and 
 in extracting collective trends.
In a previous work \cite{lambii}, we have shown that such collective listening habits may lead to the usual  music subdivisions in some particular cases, but also to unexpected structures
 that do not fit  the neat usual genres defined by the music industry. Those
 represent the non-conventional taste of listeners. Let us note that alternative music classifications based on signal analysis may also be considered  \cite{boon,ivan}.

In section II, we describe the methodology, namely the analysis of empirical data  from collaborative filtering websites, e.g. {\em audioscrobbler.com} and {\em musicmobs.com}. 
We will also give a short review of the statistical methods introduced in \cite{lambii}. Mainly, these methods consist in evaluating the correlations between the groups, depending on their audience, and in using filtering methods, i.e. percolation idea-based (PIB) methods, in order to visualise the collective behaviours.  
In section III, we attribute lists of genres to a sample of music groups, by downloading data from the web. These data, that describe the different tags, i.e. genres, used by people to classify music  groups, are analysed by using the Shannon entropy as a measure of the music group diversity.  By examining correlations between these different music genres, we also use the statistical methods of section II in order to
make a map of music genres (see \cite{cuba} for an example from the social science). This cartography is justified by the fact that {\em alike} music genres are statistically correlated by their audience. It is shown that these correlations are {\em homophilic} \cite{masuda}, i.e. {\em alike} music genres tend to be listened to by the same audience.
Homophily is known to occur in many {\em social} systems, including online communities \cite{adamic}, co-authorship networks \cite{newman0, lambiAuthors} and linking patterns between political bloggers \cite{adamic2}.

Let us stress that the issues of this work are part of the intense ongoing physicist research activity
on opinion formation \cite{holyst,galam,sznajd,castellano,kuperman}, self organisation on networks \cite{xxz,xzz}, including clique formation \cite{neil}, percolation transitions \cite{rodgers},
as well as on
 the identification of \textit{a priori} unknown collective behaviours  in complex networks \cite{vicsek}, e.g. proteins
\cite{ravasz}, genes \cite{jelena}, linguistics \cite{mike, stauf}, industrial sectors \cite{onnela}, groups of people \cite{watts}...

\section{Methodology}
\subsection{Data analysis}

In this work, we analyze data retrieved from collaborative filtering websites (see \cite{wiki} for a detailed definition). These sites  propose people to share their profiles and experiences
in order to help them discover new musics/books... that should (statistically) correspond to their own taste.
In the present case, we focus on a database downloaded from {\em audioscrobbler.com} in January 2005. It consists of a listing of  users (each represented by a number), together with the list of music groups the users own in their library. This structure directly leads to a bipartite network for the whole system. Namely, it is  a network composed by two kinds of nodes, i.e. the persons, called users or listeners in the following,  and the music groups. The network can be represented by a graph with edges running between a group $i$ and a user $\mu$, if $\mu$ owns $i$.  

\begin{figure}
\includegraphics[width=3.6in]{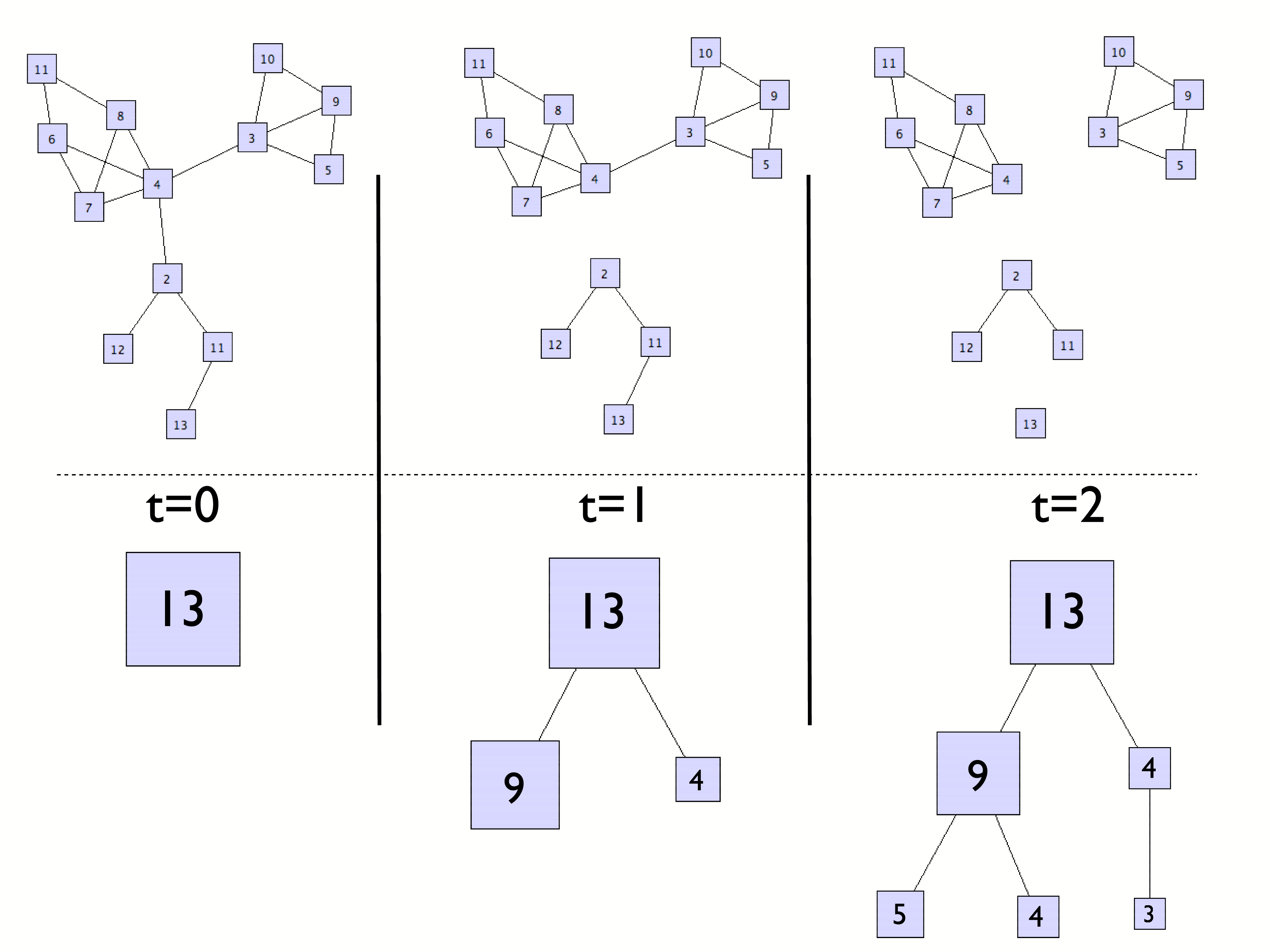}
\caption{\label{explication}  Branching representation of a squared correlation matrix of 13 elements. At each increasing step (t=0,1,2) of the filter $\phi$, links are removed, so that the network decomposes into isolated islands. These islands are represented by  squares, whose size depends on the number of nodes in the island. Islands composed by only one music group are not depicted. Starting from the largest island, branches indicate a parent relation between the islands. The increasing filter method is applied until all links are removed.}
\end{figure}

In the original data set, there are  $617900$ different music groups, although this value is skewed due to multiple (even erroneous) ways for a user to characterise an artist (e.g. {\em The Beatles}, {\em Beatles} and {\em The Beetles} count as three music groups) and $35916$ users.  On average, each user owns 140 music groups in his/her library, while each group is owned by 8 persons. 
For completeness, let us note that the listener with the most groups possesses 4072 groups ($0.6 \%$ of the total music library) while the group with the largest audience, {\em Radiohead}, has 10194 users ($28 \%$ of the user community). This asymmetry in the bipartite network is expected as users have in general specific tastes that prevent them from listening to any kind of music, while there exist {\em mainstream} groups that are listened to by a very large audience. 
Let us stress that this asymmetry is also observable in the degree distributions for the people and for the groups. 

In the following, we make a selection in the total number of groups for computational reasons, namely we have analysed a subset composed of the top 1000 most-owned groups. This limited choice was also motivated by the possibility to identify these groups at first sight.

\begin{figure}
\includegraphics[angle=-90,width=3.6in]{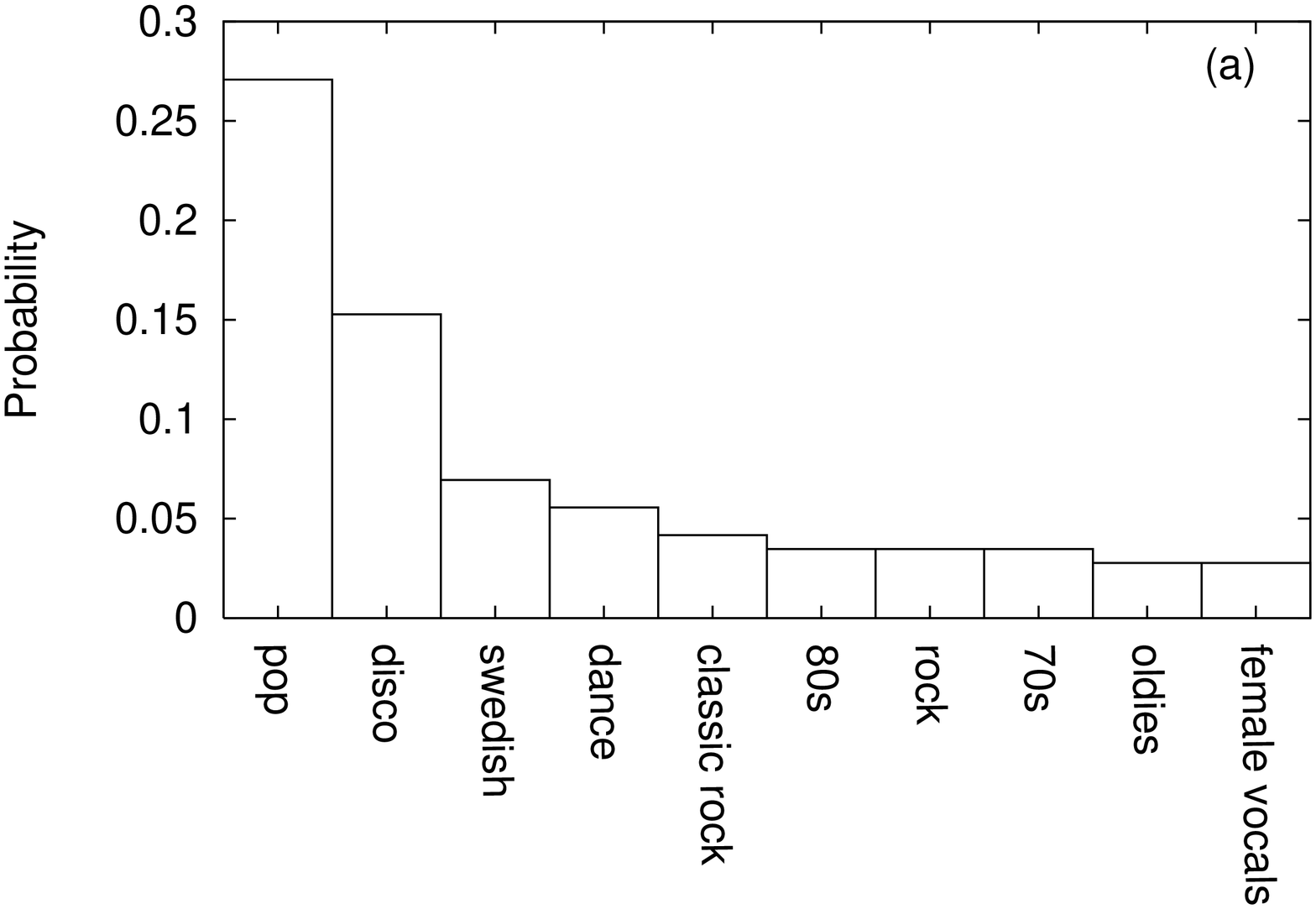}

\vspace{-1cm}
\includegraphics[angle=-90,width=3.6in]{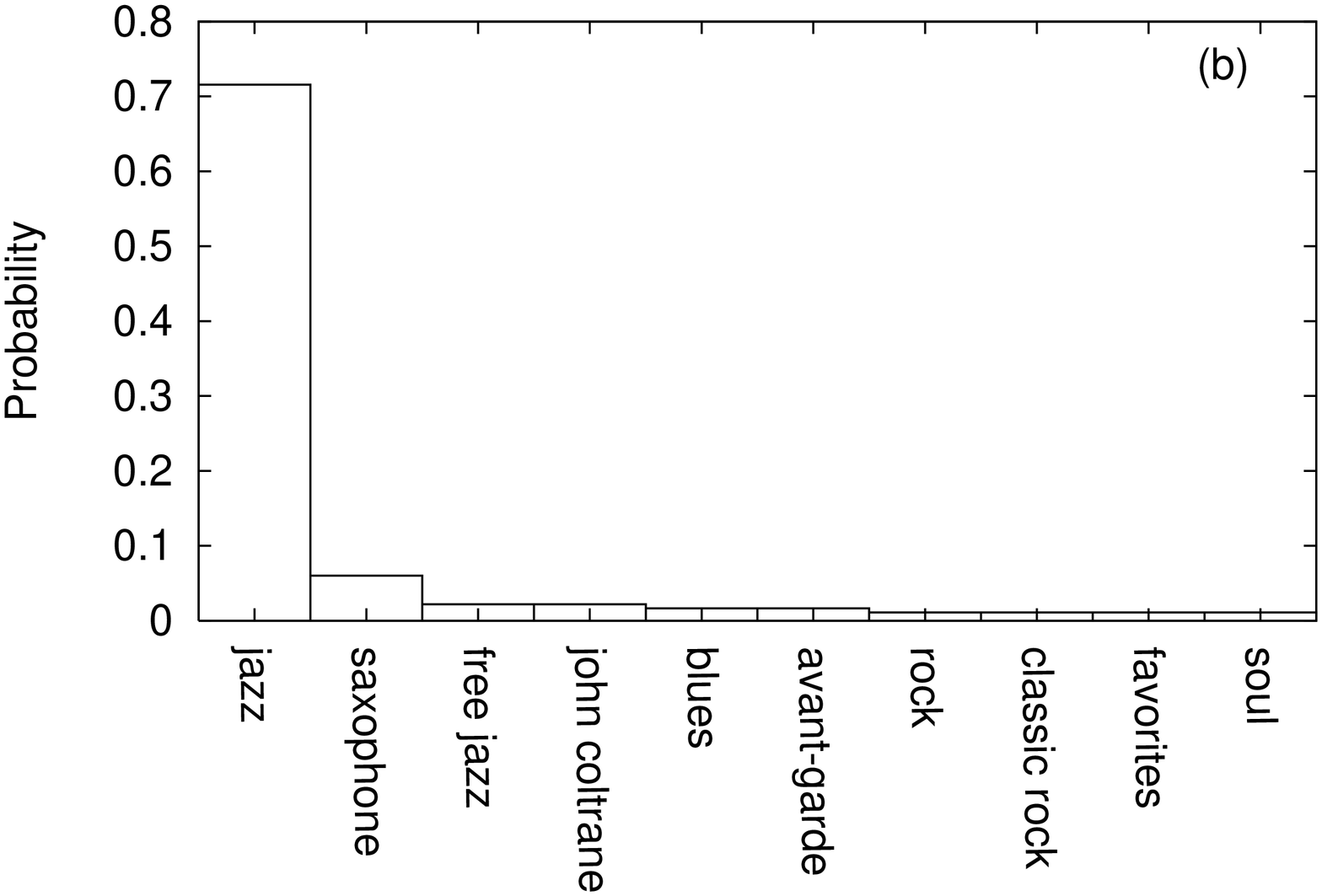}

\vspace{-1cm}
\caption{\label{abba}  Empirical probability histogram of the top 10 genres tagged by listeners to {\em ABBA} (a), and to {\em John Coltrane} (b). The data have been downloaded from {\em http://www.lastfm.com} in August 2005.}
\end{figure}

\subsection{Percolation idea-based filtering}
In this section, we review the method introduced in \cite{lambii} in order to extract collective structures from the data set.
Each music group $i$ is characterised by its  signature, i.e. the vector:
 \begin{equation}
\overline{\Gamma}^i = (..., 1, ... , 0, ... ,1, ...)
\end{equation}
of $n_L$ components, where $n_L=35916$ is the total number of users in the system, and where $\Gamma^i_{\mu}=1$ if the listener $\mu$ owns group $i$ and $\Gamma^i_{\mu}=0$ otherwise. By doing so, we consider that the audience of a music group, i.e. the list of persons listening to it,
 identifies its signature.

  In order to quantify the correlations between two music groups $i$ and $j$, we calculate the symmetric correlation measure:
\begin{equation}
\label{cosine}
C^{ij} = \frac{\overline{\Gamma}^i . \overline{\Gamma}^j}{|\overline{\Gamma}^i| |\overline{\Gamma}^i|} \equiv \cos \theta_{ij} 
\end{equation}
where $\overline{\Gamma}^i . \overline{\Gamma}^j$ denotes the scalar product between the two $n_L$-vector, and $||$ its  associated norm. This correlation measure, that corresponds to the cosine of the two vectors in the  $n_L$-dimensional space, vanishes when the groups are owned by disconnected audiences, and is equal to $1$ when their audiences strictly  overlap.

In order to extract families of alike music groups from the  correlation matrix $C^{ij}$, we use the PIB method \cite{lambii}. We define the filter coefficient $\phi \in [0,1[$, and filter the matrix elements so that $C_{\phi}^{ij}=1$ if $C^{ij}> \phi$, and let $C_{\phi}^{ij}=0$ otherwise. 
Starting from $\phi=0.0$, namely a fully connected network, increasing values of the filtering coefficient remove less correlated links and lead to the shaping of well-defined islands, completely disconnected from the main island. 
Let us stress that this systematic removal of links is directly related to percolation theory, and that the internal correlations in the network displace and broaden the percolation transition  \cite{lambii,lambi}. 
From a statistical physics point of view, the meaning of $\phi$ is that of the inverse of a temperature, i.e. 
high values of $\phi$ restrain the system to highly correlated islands; in the same way, low temperature restrains phase space exploration to low lying free energy wells.  This observation suggests that PIB methods should be helpful in visualising 
 free energy profiles and  reaction coordinates between metastable states \cite{titus}.

A branching representation of the community structuring is used to visualise the process (see Fig.\ref{explication} for the sketch of three first steps of an arbitrary example). To do so, we start the procedure with the lowest value of $\phi=0.0$, and we represent each isolated island by a square whose surface is proportional to its number of  nodes (the music groups). Then, we increase slightly the value of $\phi$, e.g. by 0.01, and we repeat the procedure.  From one step to the next step, we draw a bond between emerging sub-islands and their parent island. The filter is increased until all bonds between nodes are eroded (that is, there is only one node left in each island). Let us note that islands composed by only one music group are not depicted, as these {\em lonely} music groups are self-excluded from the network structure, whence from any genre. Applied to the above correlation matrix $C^{ij}$, the  tree structure gives some insight into the diversification process by following branches from their source (top of the figure) toward their extremity (bottom of the figure).
 The longer  a given branch is followed, the more likely it is forming a well-defined music genre.

In \cite{lambii}, we have shown that
the resulting tree representation exhibits long persisting branches, some of them leading to standard, homogenous style groupings, such as [Kylie Minogue, Dannii Minogue, Sophie Ellis Bextor] (dance pop), while many other islands are harder to explain from a standard genre-fication point of view and
reveal evidence of unexpected collective listening habits.

\section{Genre cartography}

\subsection{Measure of diversity}

\begin{figure}
\includegraphics[angle=-90,width=3.6in]{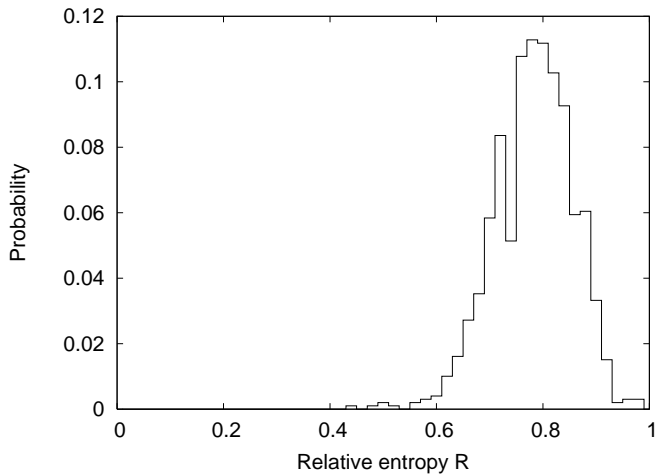}

\caption{\label{relative}  Empirical probability histogram of the relative entropy $R_i$ (see text for definition), obtained for the top 1000 music groups. The tagged genres have been downloaded from {\em http://www.lastfm.com} in August 2005.}
\end{figure}

\begin{figure*}

\includegraphics[width=2.3in]{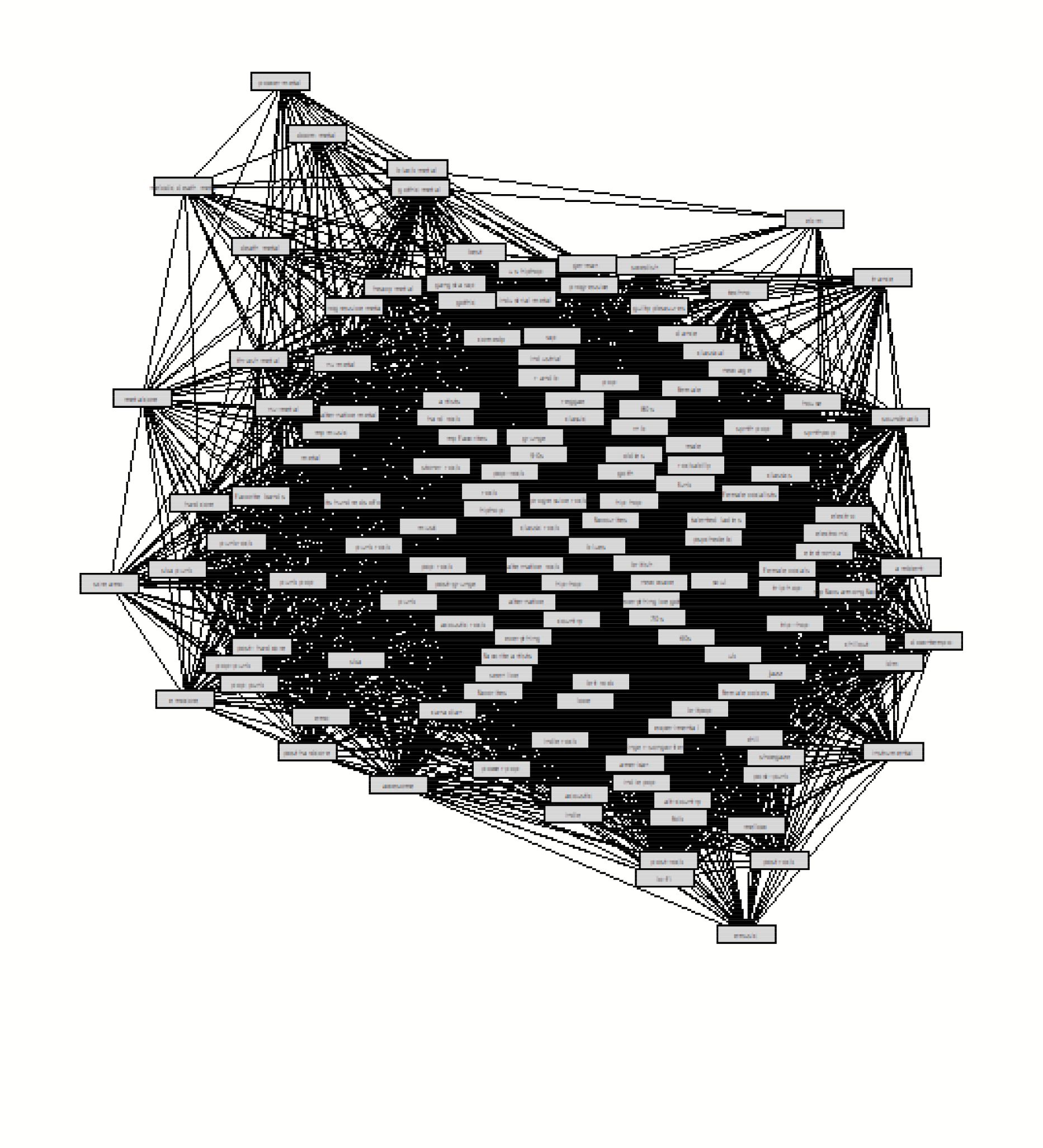}
\includegraphics[width=2.3in]{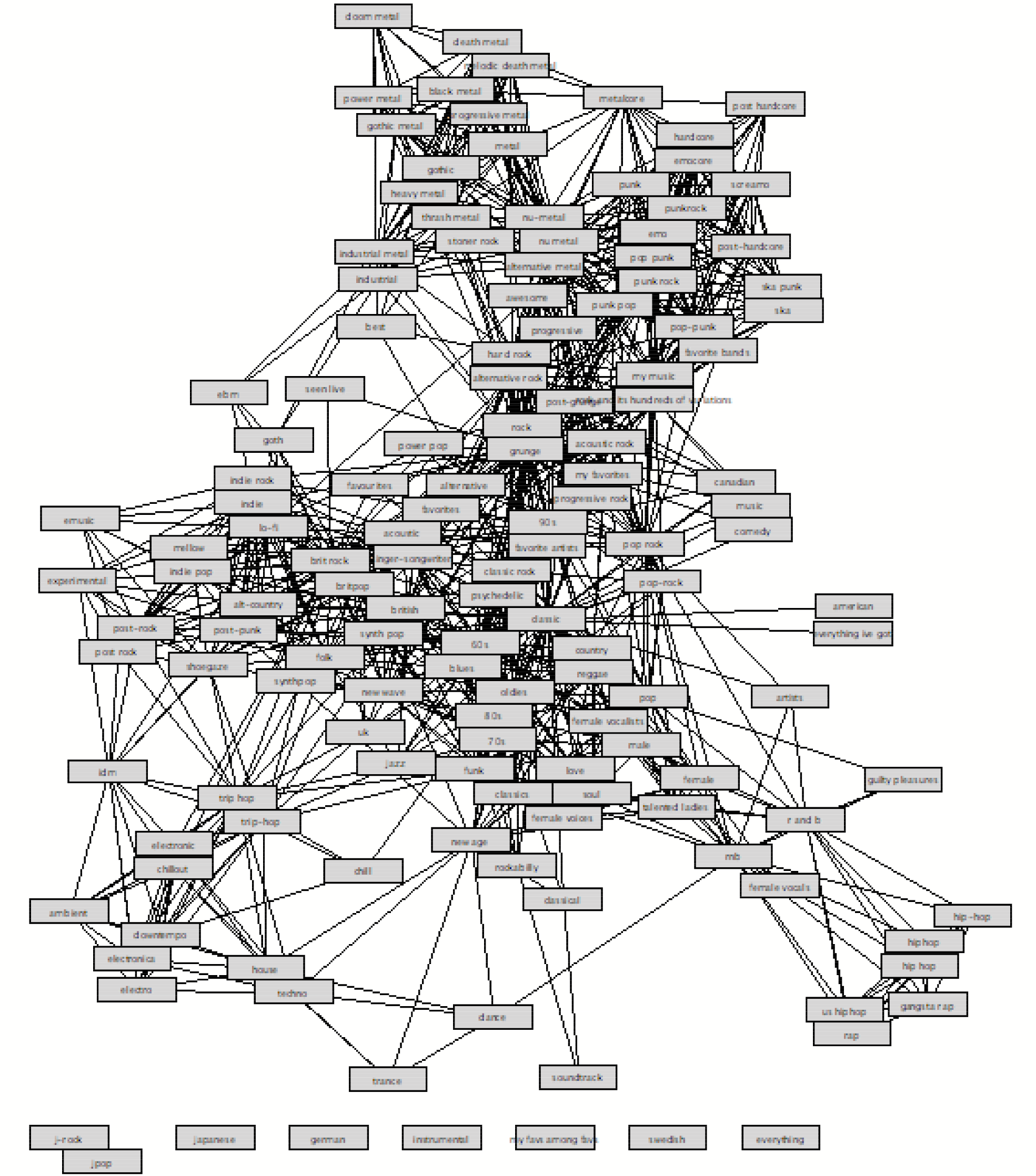}
\includegraphics[width=2.3in]{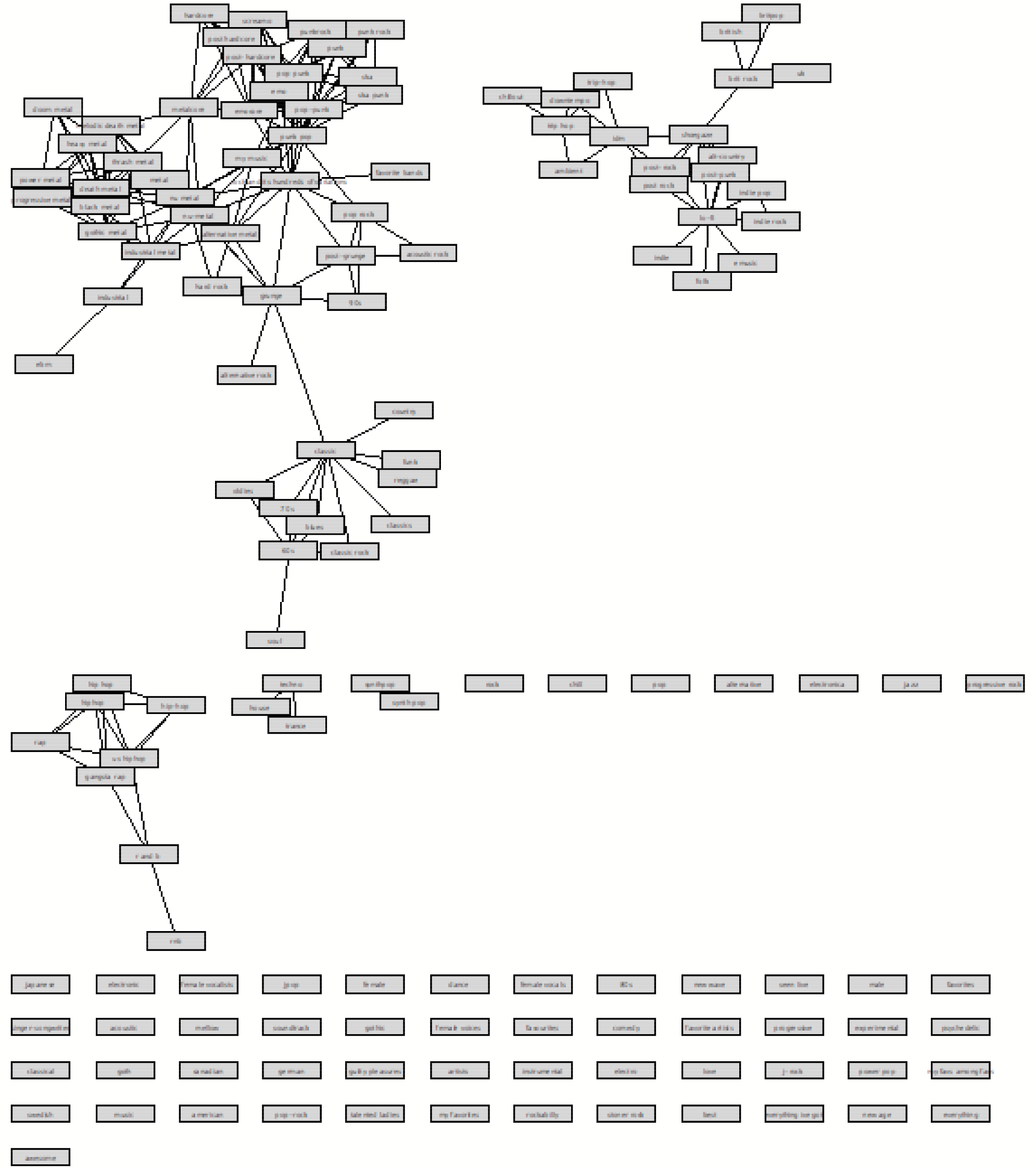}
\caption{\label{filter}  Graph representation of the music genres filtered correlation matrix $M^{\mu_1 \mu_2}$ for 3 values of the filter parameter $\phi=0.09$, $\phi=0.12$ and $\phi=0.15$, displayed from left to right. Rectangles represent the genres observed in the sample of 995 music groups. The action of filtering leads to a removal of less correlated links, thereby exhibiting the internal structure of the network.  The graphs were plotted thanks to the {\em visone} graphical tools \cite{visone}.}
\end{figure*}

In view of the above analysis, attributing genres to music groups is a difficult problem.
This complexity is made clearer by observing the different ways  listeners characterise the same music group. 
To perform this analysis, we have downloaded from $http://www.lastfm.com$ a list of the descriptions, i.e. genres, that people tag to music groups in their music library, together with the number of times this description occurred. For instance, from this site, one gets that {\em ABBA} (Fig.\ref{abba}a) is described by an eclectic range of different music sub-divisions. These sub-divisions are based on the group style (Pop, Rock...), on the time period (80s, 70s...) or on geographical grounds (Swedish) and their choice depends on the listener, i.e. his perception and subjective way to characterise music (see first paragraph of the introduction).

For this work, we have downloaded these lists of genres for the top 1000 groups, thereby empirically collecting a statistical genre-fication of the music groups. Let us stress that the data could not be downloaded for 5 of the groups, due to misprints in their name, e.g. {\em Bjadrk} instead of {\em Bj\"ork}. 
Consequently, we focus in the following on the $n_G=995$ remaining music groups.
One should also note that $http://www.lastfm.com$ limits access to the top 25 genres of each group. 

The statistical genre-fication of the sample may exhibit quantitatively different behaviours. For instance, a music group like {\em John Coltrane} (Fig.\ref{abba}b) shows a peaked histogram, i.e. it is almost only described by the tag {\em jazz}, in contrast with {ABBA} that  is described by a large variety of tags.
In order to measure the complexity, or diversity of each music group $i$, we introduce the Shannon entropy \cite{beck5}:

\begin{equation}
S_i= - \sum_i p_{i;g} \ln p_{i;g}
\end{equation}
where $p_{i;g}$ is the probability for genre $g$ to be tagged to the music group $i$, and the sum is performed over all possible genres (with, as said before, a maximum of 25). By construction, this quantity vanishes $S_i^{min}= 0$ when the group $i$ is wholly described by one tag $g^{*}$, i.e. $p_{i;g}=\delta_{gg^{*}}$ while it takes its maximum value $S_i^{max}= \ln 25$ for the uniform distribution $p_{i;g}=\frac{1}{25}$. In order to restrain the problem to the interval $[0:1]$, we introduce the relative quantity $R_i = \frac{S_i}{\ln 25}$. This quantity is therefore representative of the number of different terms needed by listeners to describe the music group $i$, i.e. the diversity of the music group. In figure \ref{relative}, we plot the empirical distribution of this relative entropy over the $995$ considered groups. It shows clearly a high degree of diversity of the music groups, therefore requesting many different tags for characterisation.

\begin{figure*}
\includegraphics[width=7.10in]{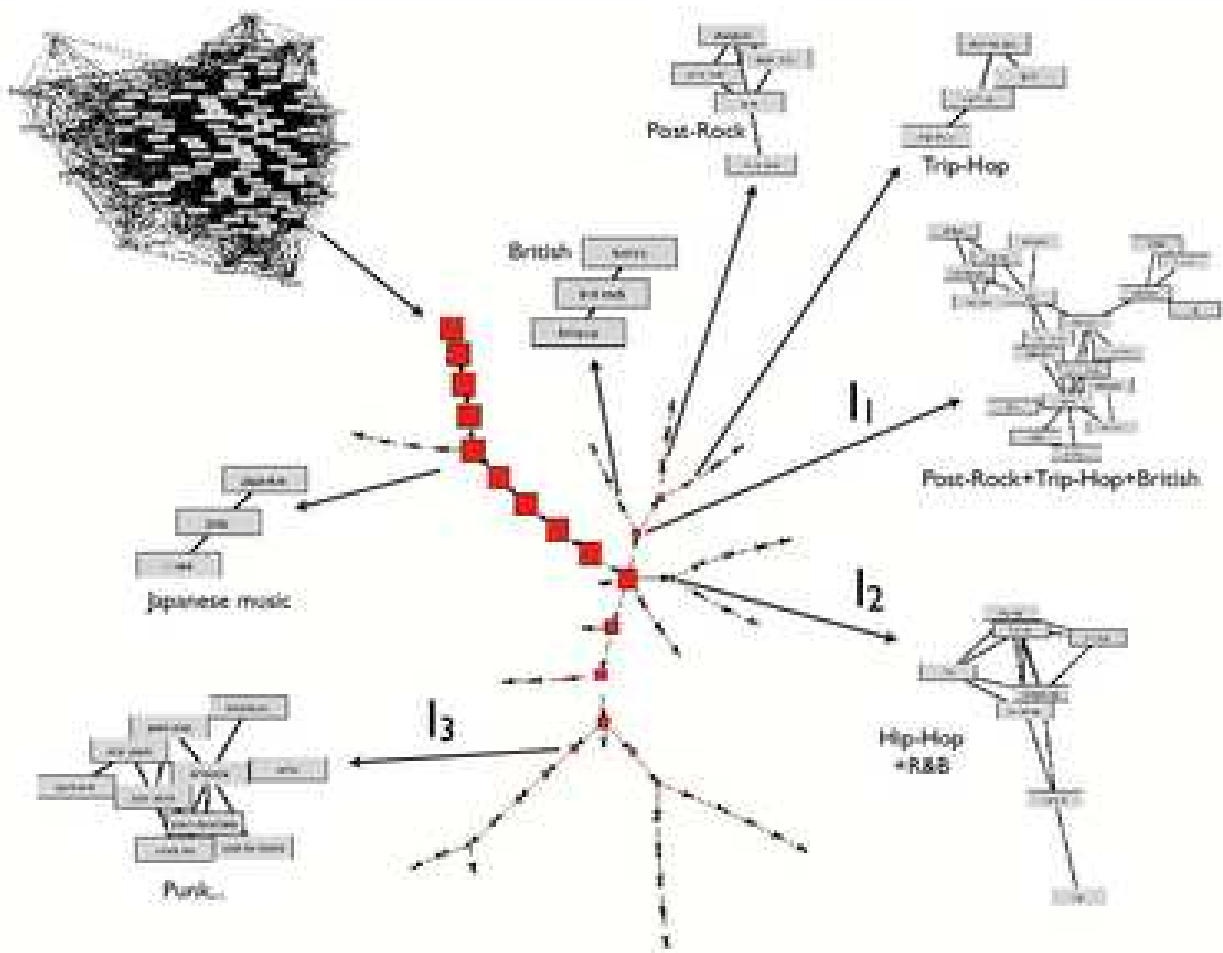}

\caption{\label{tree}  Branching representation of the correlation matrix $M^{\gamma_1 \gamma_2}$. The filtering  parameter $\phi$ ranges from $0.05$ to $0.25$ (from top to bottom), and is increased at each step by $0.01$ (the tree length is 20 steps) . It induces a snake of squares at each filtering level. The shape of the snake as well as its direction are irrelevant. The tree obviously shows the emergence of homogeneous branches, that are composed of alike music-subdivisions, thereby showing evidence of genre families. The first island extraction occurs at $\phi=0.1$, and corresponds to a family of genres related to Japanese music: {\em [japanese, jpop, j-rock]}. Among the different structures uncovered by the method, let us note the appearance of the islands $I_1$ ($\phi=0.15$), $I_2$ ($\phi=0.15$) and $I_3$ ($\phi=0.18$) described in the main text.}
\end{figure*}

\subsection{Genres correlations}

In this section, we use the methods of section IIB in order to 
 analyse the correlations between genres attributed to each music group $i$.
 In the data set, we find  $2394$ different music genres. Nonetheless,  
in order to remove irrelevant tags (due to to misprints for instance) and to simplify
our analysis, we restrict
the scope to all music genres that have been attributed to at least 20 music groups. 
There are 142 such music genres, that we label with index $\gamma \in [1,142]$.
  Let us note $G_{\gamma}$ this list of genres, and  $P_{i;\gamma}$ their probability for the music group $i$. For instance, these notations read as follows in the case of {\em John Coltrane}:

 \begin{eqnarray}
G &=& [...,jazz, ..., saxophone, ..., free~jazz, ...] \cr
P_{J.C.} &=& [..., 0.72, ..., 0.06, ..., 0.02, ...]
 \end{eqnarray}

In order to measure correlations between the 142 music genres, we define the $142 \times 142 $ correlation matrix ${\bf M}$, based on the correlations $\bf C$ between the music groups (see Eq.2):
\begin{equation}
M^{\gamma_1 \gamma_2} =  \frac{\sum_{i} \sum_{j\neq i} C^{ij} P_{i;\gamma_1}P_{j;\gamma_2}}{N^{\gamma_1 \gamma_2}}
\end{equation}
where ${\bf N}$ is a normalisation matrix:
\begin{equation}
N^{\gamma_1 \gamma_2} =  \sum_{i} \sum_{j\neq i}  P_{i;\gamma_1}P_{j;\gamma_2}\end{equation}
Practically,  we make a loop over the $n_G (n_G-1)$ pairs of {\em different} music groups $(i,j)$, each pair being characterised by the correlation coefficient $C^{ij}$.  For each of these pairs, we evaluate all pairs of music genres $\gamma_1$ and $\gamma_2$ such that $P_{i;\gamma_1}\neq 0$ and $P_{j;\gamma_2}\neq 0$, and increase the  matrix element $M^{\gamma_1 \gamma_2}$ by the quantity $C^{ij} P_{i;\gamma_1}P_{j;\gamma_2}$. The normalisation matrix element $N^{\gamma_1 \gamma_2}$ is itself updated by $P_{i;\gamma_1}P_{j;\gamma_2}$. At the end of the loops,  the correlation matrix is normalised: $M^{\gamma_1 \gamma_2} \rightarrow \frac{M^{\gamma_1 \gamma_2}}{N^{\gamma_1 \gamma_2}}$.

In order to reveal collective behaviours from the correlation matrix $M^{\gamma_1 \gamma_2}$, we apply PIB methods. Starting at a very low value of the filtering coefficient (see Fig.\ref{filter}), say $\phi=0.09$, the graph is fully connected. Increasing values of the filtering coefficient  lead to the formation of cliques and to the emergence of disconnected islands, as those occurring in \cite{lambii}. Finally, we plot in Fig.\ref{tree} the tree representation of the filtering process.
Poring over the branches of this tree is very instructive and confirms the existence of non-trivial correlations between the different music genres. These correlations shape the relations between  genres, and give an objective definition to the notion of sub-genre, genre family.... 

For instance (see Fig.\ref{tree}), one observes at $\phi=0.15$ the extraction of two large sub-islands, $I_1$ and $I_2$. $I_1$ is composed of genres related to Post-Rock, Brit-Rock and Trip-Hop: {\em [chillout, ambient, trip hop, downtempo, trip-hop, idm, post-rock, post rock, shoegaze, alt-country, post-punk, indie pop, indie rock, lo-fi, emusic, indie, folk, brit rock, british, britpop, uk]}. $I_2$ is itself composed of Hip-Hop and R\&B genres: {\em [hip hop, hiphop, hip-hop, gangsta rap, rap, us hiphop, r and b, rmb]}. At $\phi=0.16$, a small sub-island extracts from $I_1$, composed of all British related tags, thereby defining a new sub-genre. Finally, at $\phi=0.17$, $I_1$ breaks into two separated blocks, one related to Rock music, the other related to Trip-Hop music. Such a breaking also occurs for $I_2$ at $\phi=0.16$, and leads to a Hip-Hop sub-genre and a R\&B sub-genre. Finally, let us note the punk-related island $I_3$ emerging at $\phi=0.18$.

Before concluding, one should insist on the {\em homogeneity} of the above sub-islands, i.e. their composition is rational given our a priori knowledge of music.  This feature highlights the  
{\em homophily} \cite{masuda} of the music groups, which means that similar groups, i..e. groups with similar tags, tend to be listened by the same audience.

\section{Conclusion}

In this article, we study empirically the musical behaviours of a large sample of persons. Our analysis is based on web-downloaded data and uses complex network techniques in order to uncover collective trends from the data. To do so, we use  percolation idea-based techniques \cite{lambii} that consist in filtering correlation matrices, i.e. correlations between the music groups, and in visualising the resulting structures  by a branching representation.  Each of the music groups is characterised by a list of genres, that are tags used by the listeners in order to describe the music group. By studying correlations between these tags, we highlight non-trivial relations between the music genres. As a result, we draw a cartography of music, where large structures are statically uncovered and identified as a genre family.
Let us stress that this work is closely related to the theory of hidden variables \cite{agatka, boguna}, i.e. the hidden variables being here the music group tags. Consequently, this study
should provide an empirical test for the theory.

This work has also many applications in marketing and show business, e.g. taste suggestions
in online services, in publicity, libraries.... 
This kind of approach also opens the way to quantitative
modelling of opinion/taste formation \cite{lambi3}, and offers quantitative tools for sociologists and
musicologists. For instance, G. d'Arcangelo \cite{gideon2} has recently used our
analysis in order to discuss the emergence of a growing eclecticism of music
listeners that is driven by curiosity and self-identification, in opposition to the uniform trends promoted by commercial radios and $Major$ record labels \cite{margolis}. 
Applications should also be considered in taxonomy \cite{taxonomy}, in scientometrics, i.e. how to classify scientific papers depending on their authors, journal, year, keywords..., and in linguistics \cite{lingui}, in order to  highlight relations between a signifier (tag) and a signified (music group).

\begin{acknowledgments}
Figures  4 and 5 were plotted thanks to the {\em visone} graphical tools \cite{visone}.
R.L. would like to thank especially   G. D'Arcangelo for fruitful discussions. This work 
has been supported by European Commission Project 
CREEN FP6-2003-NEST-Path-012864 and COST P10 (Physics of Risks).
\end{acknowledgments}

\end{document}